\documentclass[a4paper,twocolumn,citeautoscript,prl]{revtex4-1}

\usepackage{amsmath,bm}
\usepackage[dvips]{graphicx,color}

\setcitestyle{super}

\begin{document}

\title{Synchronization of coupled optical microresonators}

\author{Jae K. Jang$^{1}$, Alexander Klenner$^{1}$, Xingchen Ji$^{2,3}$, Yoshitomo Okawachi$^{1}$, \\ Michal Lipson$^{3}$, and Alexander L. Gaeta$^{1,*}$}

\affiliation{$^{1}$Department~of~Applied~Physics~and~Applied~Mathematics,~Columbia~University,~New York,~NY~10027\\
             $^{2}$School of Electrical and Computer Engineering, Cornell University, Ithaca, NY 14853\\
             $^{3}$Department of Electrical Engineering, Columbia University, New York, NY 10027}


\begin{abstract}
  \noindent The phenomenon of synchronization occurs universally across the natural sciences and provides critical insight into the behavior of coupled nonlinear dynamical systems. It also offers a powerful approach to robust frequency or temporal locking in diverse applications including communications, superconductors, and photonics. Here we report the experimental synchronization of two coupled soliton modelocked chip-based frequency combs separated over distances of 20~m. We show that such a system obeys the universal Kuramoto model for synchronization and that the cavity solitons from the microresonators can be coherently combined which overcomes the fundamental power limit of microresonator-based combs. This study could significantly expand applications of microresonator combs, and with its capability for massive integration, offers a chip-based photonic platform for exploring complex nonlinear systems.
\end{abstract}

\maketitle

\noindent Synchronization behavior is ubiquitous in nature~\cite{strogatz_from_2000} and has been studied in a vast variety of systems, such as synchronously flashing fireflies~\cite{buck_synchronous_1988}, pacemaker cells in mammalian hearts~\cite{michaels_mechanisms_1987}, superconducting Josephson junctions~\cite{wiesenfeld_synchronization_1996}, and a network of microwave oscillators~\cite{york_quasi_1991}. Interestingly, the first scientific observation of synchronization dates back to the 17th century when Christiaan Huygens discovered that the periods of two pendulum clocks hanging on a common wooden beam tend to lock~\cite{ramirez_the_2016}. He called this observation ``odd sympathy'' and suspected that mechanical coupling was responsible for it. Over the course of subsequent studies, it has been confirmed and is now widely accepted that coupling among the constituents of a system plays a key role in their mutual synchronization. This conceptual understanding has been incorporated into several mathematical models~\cite{strogatz_from_2000}, most notably the Kuramoto model~\cite{kuramoto_self_1975}, with remarkable success and universal applicability. Only in recent years has this concept been extended to optical systems, such as networks of coupled monochromatic lasers, leading to demonstration of phase-locking~\cite{nixon_controlling_2012} and coherent beam combining~\cite{cheo_a_2001}.

The optical frequency comb represents an electromagnetic excitation comprised of a discrete set of equidistant spectral lines and emerged as an important subfield of optics with numerous applications ranging from spectroscopy to ultrafast optics and metrology~\cite{udem_optical_2002,cundiff_colloquium_2003,dudley_supercontinuum_2006,newbury_searching_2011}. Such a source establishes a link between the time-domain description of a uniform pulse train, as might be generated by a modelocked laser, and the associated ``picket-fence'' frequency-domain picture of the laser output spectrum. It is within this context where the possibility of achieving a timing correlation between two modelocked lasers was initially investigated~\cite{cundiff_colloquium_2003}. More recently, it has been shown that a microresonator driven by an external single-frequency pump field is also capable of generating a frequency comb~\cite{delhaye_optical_2007,savchenkov_tunable_2008,levy_cmos_2010,razzari_cmos_2010}. The microresonator sustains its comb through nonlinear parametric interactions driven by the external pump and exhibits behavior that is substantially different to that of modelocked lasers~\cite{herr_universal_2012,leo_dynamics_2013,parrarivas_dynamics_2014,anderson_observations_2016,bao_observation_2016,yu_breather_2017,
lucas_breathing_2017}. Such microresonator-based frequency combs can be modelocked by the excitation of intracavity dissipative solitary pulses known as temporal cavity solitons~(TCS's)~\cite{leo_temporal_2010,saha_modelocking_2013,
herr_temporal_2014,jang_temporal_2015,delhaye_phase_2015,liang_high_2015,yi_soliton_2015,joshi_thermally_2016,wang_intracavity_2016,webb_experimental_2016} and offer the merits of robustness, compactness, and potential for on-chip integration. Despite extensive studies of the dynamics of single microresonators, only recently have there been preliminary theoretical studies exploring evanescently coupled microresonators~\cite{wen_synchronization_2015,munns_novel_2017}, which predict frequency-locking between the two generated combs.

Here, we experimentally demonstrate passive synchronization of two microresonator-based optical frequency combs. We generate modelocked combs in two silicon nitride~(Si$_3$N$_4$) microresonators on separate chips and synchronize them by coupling a small fraction of one microresonator output to the input of the other microresonator via an optical fiber link that is 20~m long. In addition, we report on successful coherent combining of the outputs of two synchronized cavity soliton combs, which could have immediate implications in overcoming the power limitation of microresonator comb technology. Using a system of coupled Lugiato-Lefever equations~\cite{coen_modeling_2013,chembo_spatiotemporal_2013,wabnitz_suppression_1993,lugiato_spatial_1987}, we find excellent agreement with our experimental results and show that such a system can be reduced to the Kuramoto model. Our demonstration offers the prospect of applications such as the synchronization of multiple wavelength-division multiplexed sources~\cite{levy_high_2012,marin-palomo_microresonator_2017}, synthetic aperture imaging and clock distribution~\cite{newbury_searching_2011}.

\section*{Illustration of concept}

\begin{figure*}[t]
\centerline{\includegraphics[width=15cm]{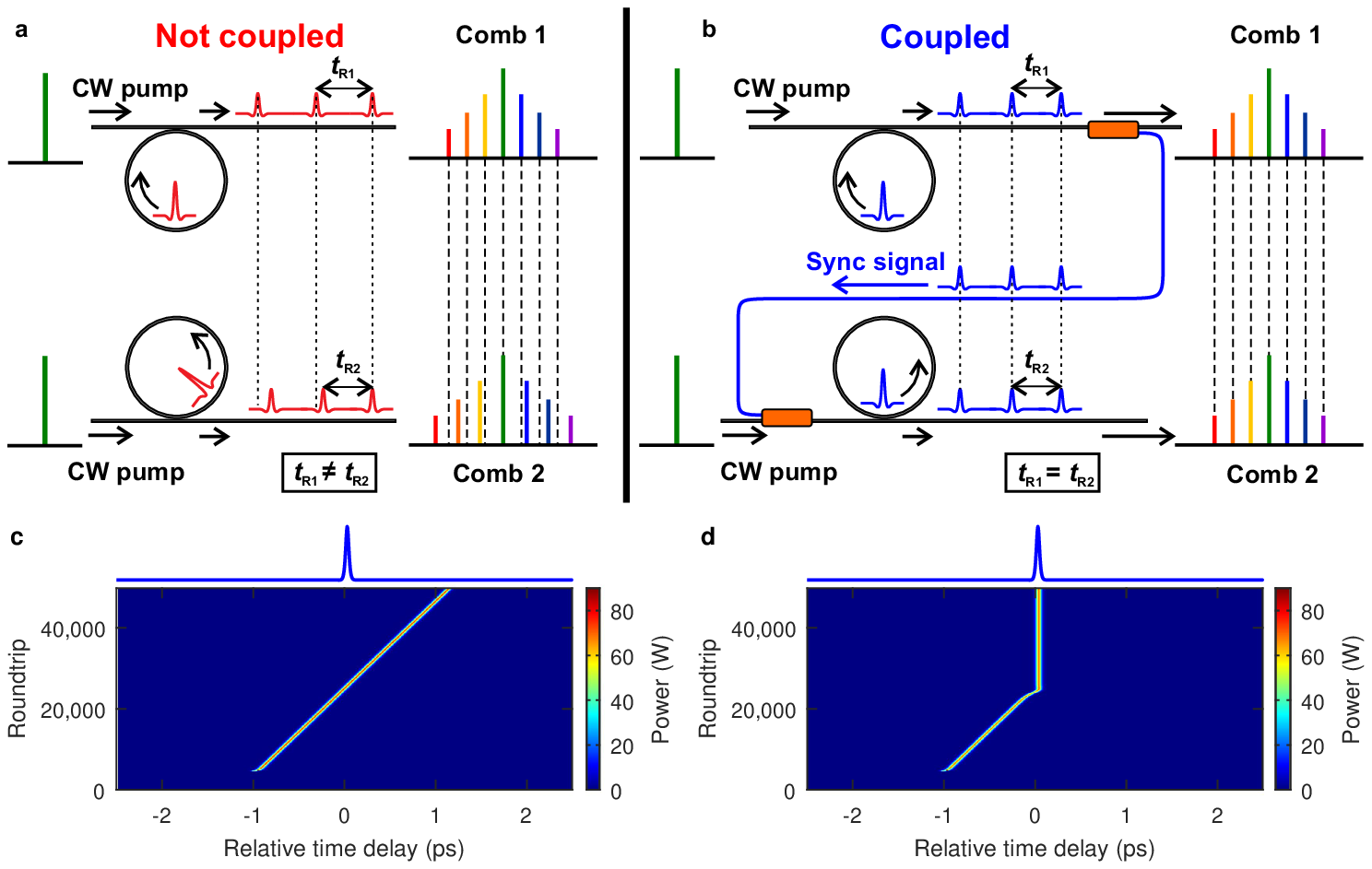}}
\caption{\textbf{Illustration of the synchronization of microresonator combs.} \textbf{a,} A schematic of the scenario without coupling. The circulating TCS pulses inside the two microresonators have different roundtrip times, $t_{\mathrm{R},1}$ and $t_{\mathrm{R},2}$, which results in the spacings of the two combs to be different. \textbf{c,} The corresponding numerical simulation shows that the pulse of the slave resonator drifts relative to the reference (master) pulse, shown as blue curve on top. \textbf{b,} When coupling is established, it is possible to synchronize the pulses and lock the frequencies of the combs such that the comb spacings become identical. \textbf{d,} Synchronization manifests as stable locking of the relative positions of the TCS's. CW: continuous-wave.}
\label{fig:fig1}
\end{figure*}

\begin{figure*}[t]
\centerline{\includegraphics[width=15cm]{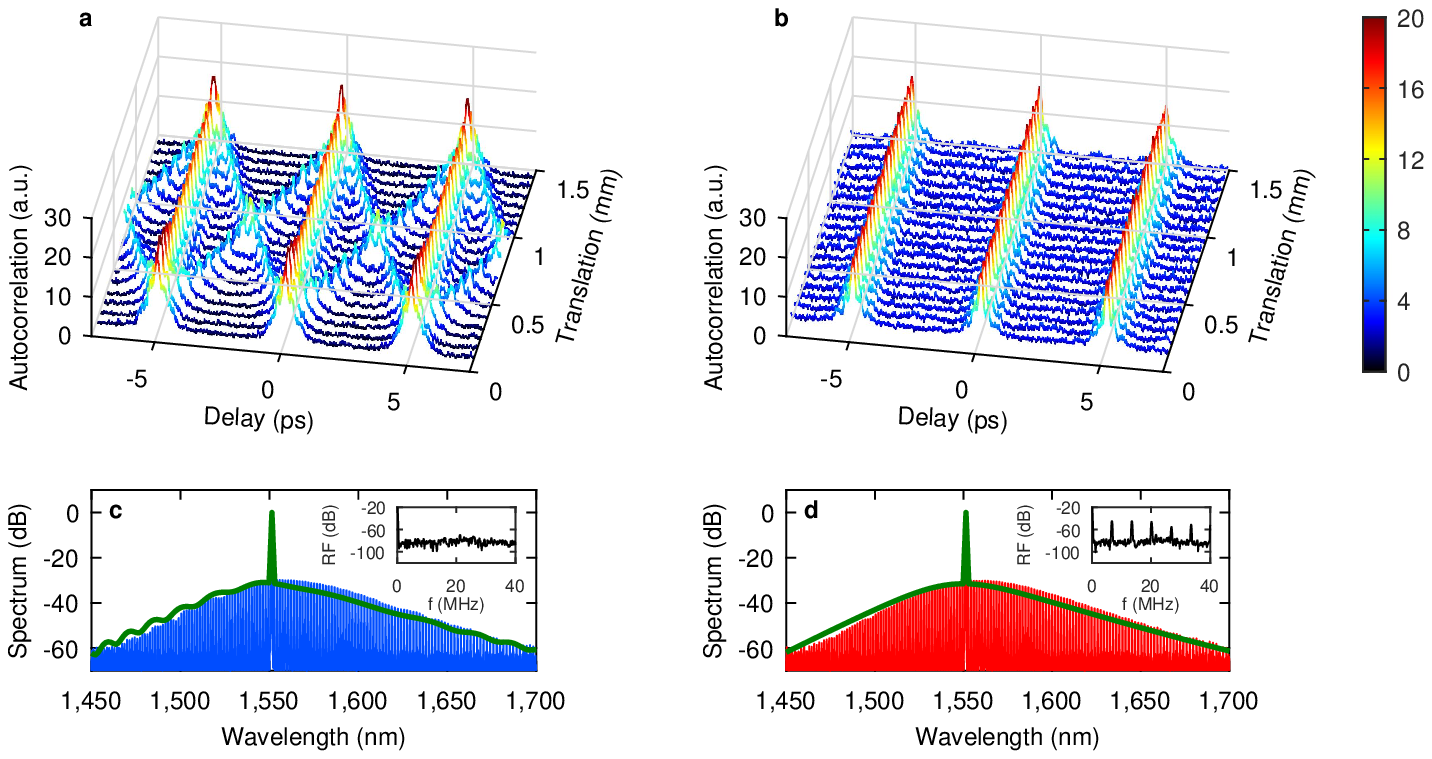}}
\caption{\textbf{Dynamics of synchronization.} \textbf{a,b} Autocorrelation traces of the combined comb signal for the synchronized (\textbf{a}) and unsynchronized cases (\textbf{b}). When the combs are synchronized, sharp secondary peaks, as well as primary peaks with a 5~ps period, are observed. These secondary peaks can be shifted as the relative delay between the two comb outputs is tuned. \textbf{c,} In this case, we also observe spectral fringes in the spectrum of the slave comb, and the absence of RF beatnotes (inset). \textbf{b,} When the combs are not synchronized, no secondary peak is observed and background level is significantly higher. \textbf{d,} In this case, the spectral fringes disappear in the optical spectrum and sharp beatnote harmonics appear (inset).}
\label{fig:fig2}
\end{figure*}

\noindent Our synchronization scheme is illustrated in Fig.~\ref{fig:fig1}. Each microresonator has an integrated microheater, which is electrically tuned to access the modelocked comb state associated with a single TCS~\cite{joshi_thermally_2016}. The heater voltage can also be adjusted after entering the modelocked state. This thermally shifts resonances~\cite{carmon_dynamical_2004} relative to the pump and enables fine tuning of the repetition rate of the circulating TCS or equivalently of the comb spacing ($\sim 200$~GHz), in each microresonator (see Methods for further information on experimental setup). In the absence of physical coupling (Fig.~\ref{fig:fig1}a), the microresonators sustain a TCS each, with roundtrip times ($t_{\mathrm{R},1}$ and $t_{\mathrm{R},2}$ which are the reciprocal of the respective repetition rates) that fluctuate relative to each other due to factors such as spatially varying environment. This mismatch in the roundtrip times is directly correlated to fluctuations in the relative spacing of the associated combs. We establish a coupling link between the microresonators by collecting the output of one microresonator and transmitting a small portion ($<1~\%$) of it through an optical fiber link to the other microresonator (Fig.~\ref{fig:fig1}b). The transmitted signal is combined with the pump and the total field drives the second resonator. The first microresonator serves as the reference and constantly transmits a fraction of its output to the other microresonator. Due to this coupling signal, the optical field inside the second microresonator develops localized intensity and phase modulations that enforce synchronization (physical mechanism is discussed in Supplementary Information; See Supplementary Figs.~S5 and 6). For this reason, we refer to the former and the latter as the master and slave microresonators, respectively, and the coupling signal as the sync signal. We elucidate the dynamics of synchronization in Figs.~\ref{fig:fig1}c and d with numerical simulation based on Lugiato-Lefever equations, details of which are described in the Methods and Supplementary Information. These density plots display the evolutions of the slave TCS without (Fig.~\ref{fig:fig1}c) and with coupling (Fig.~\ref{fig:fig1}d), with respect to the master TCS shown as blue curves on top. In both cases, we begin the simulation by seeding the master resonator with an approximate analytic TCS solution~\cite{herr_temporal_2014,wabnitz_suppression_1993}. After 5,000 roundtrips, we excite another TCS in the slave resonator. With no coupling, the slave TCS drifts relative to the master TCS at a constant rate chosen to be 2~MHz. When coupling is introduced, the slave TCS locks its position to that of the master TCS such that their roundtrip times become synchronized. Following the theoretical work of~\cite{jang_temporal_2015} and making appropriate assumptions, we derive (in Supplementary Information) the dynamical equation of the temporal position $\tau_2$ of the enslaved TCS relative to the master TCS which can be written as
\begin{equation}\label{eq:1}
\frac{d\tau_2}{dt'}=\Delta\tau-k\sin\bigg(2\pi\frac{\tau_2}{t_{\mathrm{R},1}}\bigg)
\end{equation}
where $t'$ counts the number of roundtrips, $\Delta\tau$ is a relative drift rate per roundtrip in the absence of coupling, $k$ is coupling constant that depends on the coupling link transmission, and $t_{\mathrm{R},1}$ is the roundtrip time of the master TCS. Equation~(\ref{eq:1}) is mathematically equivalent to the Kuramoto model~\cite{kuramoto_self_1975} of a system of two oscillators with unidirectional coupling. The model predicts that the equilibrium condition (\emph{i.e.} synchronization) can be realized if the coupling strength exceeds the natural drift rate $\Delta\tau$ of the slave TCS.

\section*{Results}

\begin{figure*}[t]
\centerline{\includegraphics[width=15cm]{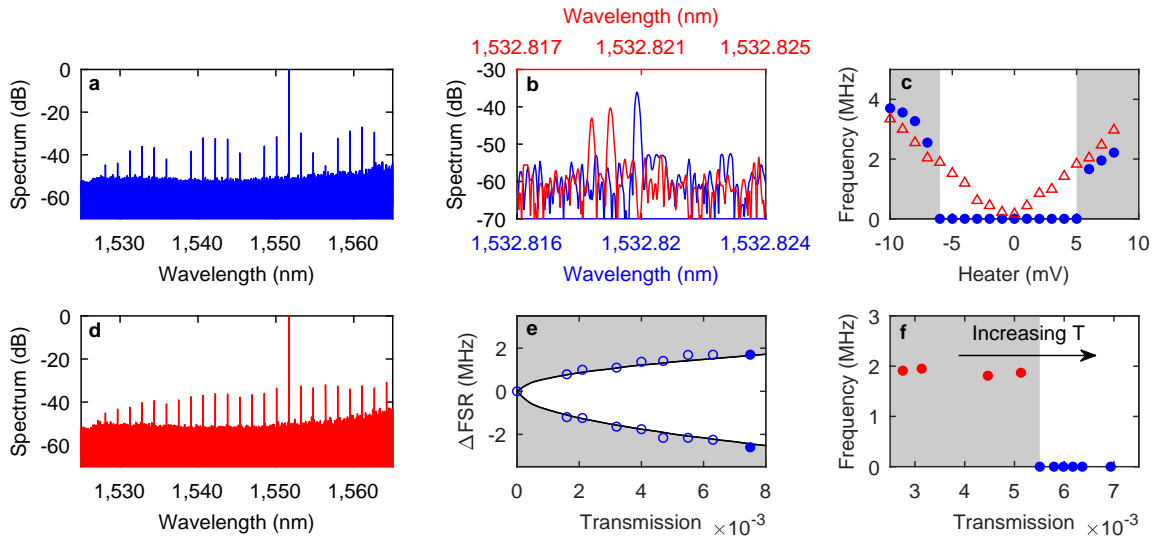}}
\caption{\textbf{Characterization of synchronization behavior.} Here the results of the synchronized case
are plotted in blue and those of the unsynchronized case in red. Grey regions indicate the
unsynchronized zones. \textbf{a,d,} High-resolution optical spectrum traces of the combined
comb for the synchronized (\textbf{a}) and the unsynchronized cases (\textbf{d}), respectively. \textbf{b,} Expanded
plots of the 12th comb line from the pump, marked by black dashed rectangles in \textbf{a} and \textbf{d}. One
of the plots has been displaced along the wavelength axis by 0.001~nm for clarity. \textbf{c,} Fundamental
beatnote frequency as the slave resonator is thermally tuned, when the combs are coupled (blue
circles) and uncoupled (red triangles). \textbf{e,} The maximum comb spacing mismatch ($\Delta FSR$) allowed for varying
transmission of the sync signal (blue circles) and the corresponding numerical prediction (black
curves). The solid circles are derived from \textbf{c}. \textbf{f,} The evolution of the beatnote as the coupling
strength is varied slowly, showing an abrupt transition to synchronization.}
\label{fig:fig3}
\end{figure*}

\begin{figure*}[t]
\centerline{\includegraphics[width=12cm]{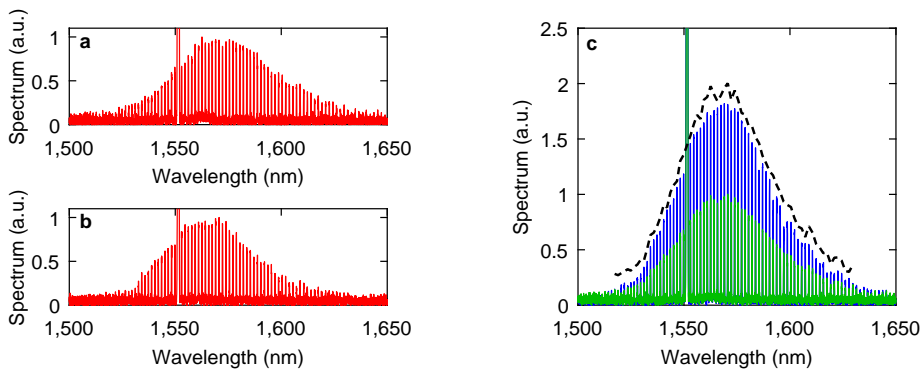}}
\caption{\textbf{Demonstration of coherent comb combining.} \textbf{a,b,} The individual comb spectra before they are combined. \textbf{c,} Blue curve displays the measured combined spectrum when the combs are synchronized and aligned in time. The black dashed line shows the theoretically anticipated level calculated from the spectra of \textbf{a} and \textbf{b}. In comparison, the green curve corresponds to the incoherently combined spectrum.}
\label{fig:fig4}
\end{figure*}

\noindent We experimentally probe the temporal dynamics of synchronization by performing an intensity autocorrelation measurement of the \emph{combined} comb signals (see Methods and Supplementary Fig.~S2). We introduce a translation stage into one of the microresonator outputs to control the relative delay between the two before they are combined. We translate the stage over a total range of 1.5~mm, corresponding to one pulse period ($\sim 5$~ps) of a 200~GHz repetition rate pulse train. Figure~\ref{fig:fig2}a shows a series of autocorrelation traces for a range of delays when the two combs are locked via the synchronization process. For each trace, a regular set of primary peaks with a 5~ps period is observed, as expected. What distinguishes these traces from those of the unsynchronized case displayed in Fig.~\ref{fig:fig2}b is the appearance of a symmetric pair of secondary peaks around each primary peak~\cite{trebino_the_2000}. Although each microresonator sustains only one TCS per roundtrip, the combined signal consists of interleaved pulse trains with a relative separation resulting from the output optical path length difference, that is, by how much extra distance the second pulse train must travel compared to the first train before they are combined. The fact that the secondary peaks are well-defined indicates that this relative separation remains stable~\cite{trebino_the_2000}. Varying the delay of one arm enables us to tune the separation. In addition, we present the optical spectrum of the slave comb in Fig.~\ref{fig:fig2}c and the complementary frequency down-converted beatnote measurement in the corresponding inset, for the synchronized case. We identify the fringes in the optical spectrum as a signature of synchronization, which we attribute to stable interference between the slave comb and sync signal. Since the fringe spacing is determined by the temporal separation between the comb and the sync signal, such a stationary interference can be observed only if the timing of the circulating TCS's in the two microresonators is synchronized. The absence of beatnotes also implies that the comb lines are spectrally aligned and that the spacings of the combs are identical. The result of numerical simulation, plotted as a green curve on the same panel, is in good agreement with the experiment and confirms the underlying dynamics. In the unsynchronized scenario in Fig.~\ref{fig:fig2}b, on the other hand, the secondary peaks disappear and instead, we notice a significant increase in the background level of autocorrelation. Also the fringes of the slave comb disappear, and we simultaneously observe well-defined beatnote harmonics, as displayed in Fig.~\ref{fig:fig2}d. The dynamics in this case are analogous to an unstable two-pulse regime with a roundtrip-to-roundtrip fluctuation in the relative positions of the pulses, leading to increased background level of the autocorrelation traces~\cite{trebino_the_2000}, and to the disappearance of the spectral fringes. The numerical simulation again reveals the underlying dynamics of this scenario.

We present additional experimental evidence of robust comb synchronization in Fig.~\ref{fig:fig3}. Here we measure the combined comb spectra with a high-resolution (10~MHz) optical spectrum analyzer (Figs.~\ref{fig:fig3}a and d). When the combs are synchronized, the spectrum in Fig.~\ref{fig:fig3}a exhibits high-contrast modulation and is indistinguishable from the output spectrum of a single microresonator operating in a stable two-pulse regime~\cite{joshi_thermally_2016}. Thus the outputs of the two microresonators are mutually phase-locked and are coherently combined. Moreover, when we zoom in on one of the comb lines corresponding to the $12$th line from the pump (enclosed with a dashed box in Fig.~\ref{fig:fig3}a), we observe a single comb line as shown in Fig.~\ref{fig:fig3}b (blue), which confirms that the spectral lines of the two combs are aligned and that the combs are frequency-locked. This scenario is to be contrasted with the unsynchronized case where no modulation is observed in Fig.~\ref{fig:fig3}d, which is a consequence of the drift and fluctuation of the relative intracavity TCS positions. As a result, the roundtrip averaged optical spectrum in Fig.~\ref{fig:fig3}d displays no interference pattern. In addition, the expanded plot of the $12$th spectral component in Fig.~\ref{fig:fig3}b displays two distinct lines (red), as the comb spacings are mismatched. In Fig.~\ref{fig:fig3}c, we present the observed variation in the first-harmonic beatnote frequency as the slave resonator is thermally tuned, both in the presence (blue circles) and absence (red triangles) of coupling (see also Supplementary Fig.~S4). Without coupling, the beatnote frequency decreases smoothly, crosses zero and increases. This trend is contrary to the coupled scenario displaying a total range of 11~mV of heater voltage over which no beatnote is observed (white indicates the synchronized region). Such a frequency-locking regime is similar to that reported in the context of counter-propagating solitons in a single microresonator~\cite{yang_counter_2017,joshi_counter_2018}. This range of heater voltage corresponds to 400~MHz resonance shift relative to the pump (see Methods for calibration). This result also shows that the allowed comb spacing mismatch between the two combs for this particular coupling strength is approximately 2.6~MHz or 1.7~MHz, depending on the sign of the mismatch (whether the spacing of the slave comb is larger or smaller than the master resonator). This slight asymmetry arises due to frequency-dependence of the collection optics and the fiber link (Supplementary Fig.~S3). These values are in good agreement with the predicted values of 2.5~MHz and 1.5~MHz from our simulations. We show in Fig.~\ref{fig:fig3}e that these maximum allowed spacing mismatch depends on the coupling transmission strength, where the simulation (black curves) again correctly predicts the experimental data (blue circles). Finally, we demonstrate the behavior of the beatnote in Fig.~\ref{fig:fig3}f as the coupling transmission is gradually ramped up, while both combs persist. Initially for low coupling strength, the fundamental beatnote is observed at around 2~MHz. It disappears abruptly as coupling is increased, implying the existence of critical coupling strength for synchronization, as predicted by the Kuramoto model Eq.~(\ref{eq:1}), and experimentally and numerically in Fig.~\ref{fig:fig3}e.

Figure~\ref{fig:fig3}a shows that combining two synchronized modelocked combs can result in stationary interference, which suggests coherent comb combining as a natural application of synchronization. An example of the constructive interference of the individual comb lines, leading to higher comb power is demonstrated in Fig.~\ref{fig:fig4}. We prepare the two individual combs in Figs.~\ref{fig:fig4}a and b to have comparable powers before combining them via a beam splitter. The combined comb spectrum is plotted blue in Fig.~\ref{fig:fig4}c. For comparison, we plot the incoherently combined spectrum in green on the same axis. As can be seen, the coherently combined comb has nearly double the power of its incoherent counterpart, and closely matches the expected coherently combined spectral power, theoretically calculated from the spectra in Figs.~\ref{fig:fig4}a and b. This demonstration offers a promising approach to boost the overall power of microresonator-based frequency combs. By synchronizing multiple microresonator combs with the technique presented here, it should be possible to circumvent output power limitation set by the efficiency of a single microresonator-based comb generator, which has been shown to be of the order of $1~\%$~\cite{yi_soliton_2015,wang_intracavity_2016}. As we require less than $1~\%$ of the output power of a comb to synchronize it to another, it is possible in principle to boost the power with near $100~\%$ efficiency.

\section*{Conclusion and outlook}

We have investigated synchronization of two microresonator-based modelocked frequency combs, and unveiled the underlying dynamics both in the time and frequency domains. This phenomenon can be understood in terms of coupling-induced interaction between the circulating temporal cavity solitons in the microresonators, leading to their temporal synchronization and hence, frequency-locking of the associated coherent combs. Based on this new-found knowledge, we have also demonstrated coherent combining of the two mutually synchronized comb. We emphasize that this technique can benefit considerably from the existing technology of photonic integration which offers superior performance and stability. It should be possible to create a fully integrated on-chip comb generator and combiner, consisting of multiple linked microresonators and interferometers that would enable power levels far exceeding those typical in single microresonator systems.

Microresonator combs have sparked interest in data telecommunications~\cite{levy_high_2012,marin-palomo_microresonator_2017}, where their functionalities both as wavelength-division multiplexing sources on the transmitter end and as local oscillators on the receiver end have been demonstrated~\cite{marin-palomo_microresonator_2017}. The synchronization technique studied here could further advance such a coherent communication scheme, offering a means of achieving phase coherence between the multi-carrier sources and the local oscillators. Our simulation (Supplementary Fig.~S7) reveals that it is not necessary to use the entire bandwidth of a comb as the sync signal, which implies that limited bandwidth of an optical amplifier would not be a hindrance to our scheme. In fact, the use of amplifiers can compensate for losses, which could enable applications that require long-distance synchronization, such as aperture imaging and clock distribution~\cite{newbury_searching_2011}. Furthermore, as we have shown, the time-domain dynamics of two coupled microresonators can be reduced to the universal Kuramoto model. Combined with a recent treatment of a single microresonator comb as coupled phase oscillators~\cite{wen_self_2016}, this correspondence suggests microresonator networks can serve as a novel platform for exploring the dynamics of complex systems.

\section*{Methods}

\footnotesize

\subparagraph*{Experimental setup. }
A detailed schematics of our experimental setup is presented in Supplementary Fig.~S1. We use two Si$_3$N$_4$ microring resonators with a waveguide cross-section of $730\times1500$~nm, on two independent chips. Both microresonators have a free-spectral-range of approximately 200~GHz with the corresponding resonator length of 830~$\mathrm{\mu m}$. They are fabricated using techniques similar to those reported in~\cite{ji_ultra_2017}. We also fabricate an integrated platinum microheater on top of each ring resonator which allows us to electrically tune the resonance frequency detuning relative to the pump frequency~\cite{carmon_dynamical_2004}, and to access the single cavity soliton-based coherent frequency comb state using an arbitrary waveform generator. The pump sources for both microresonators are derived from a single continuous-wave~(CW) laser (TOPTICA CTL 1550) whose output at 1551.4~nm is amplified with an erbium-doped fiber amplifier and is split into two light fields of equal intensity with a 50/50 fused fiber coupler. Each field is coupled to the integrated bus waveguide of a microresonator via a lensed fiber. The output of each microresonator is collected into an optical fiber with a microscope objective and a fiber collimator package. It is split with a 90/10 fiber coupler where the lower intensity beam is used to monitor the optical spectrum with an optical spectrum analyzer~(OSA). The remaining beam is spectrally decomposed into its pump and comb components using a fiber dense wavelength-division-multiplexer~(DWDM) with $\sim 50$~GHz passband centered at 1551.72~nm. The pump component enables us to monitor the relative transmission of the microresonator on an oscilloscope, while the comb component is used for a variety of measurements, such as beatnote measurement with an electronic spectrum analyzer. In addition, polarization beam splitters are inserted in the free-space output paths of the resonators, where small fractions ($\sim 1~\%$) of light are reflected. We coherently combine the reflected beams with a non-polarizing beam splitter and detect the combined beam with a 10-MHz resolution OSA (Aragon Photonics BOSA 400).

The comb component of one microresonator is further split into two beams with a 50/50 coupler. One of the split beams, which consists of a fraction of the collected comb signal, is transmitted through a fiber link before it is combined with the second CW pump beam through a DWDM and drives the other microresonator. We install a polarization controller in the fiber link to ensure that the polarization state of the transmitted comb signal is collinear with that of the second pump. As the fiber link consists of many fiber components whose combined length is about 15.5~m of standard telecom single-mode-fiber (SMF) with anomalous group-velocity dispersion (GVD) of $\beta_{2,\mathrm{SMF}} = -21.4$~$\mathrm{ps^2/km}$ at 1551.4~nm, we add 6.8~m of dispersion compensating fiber (DCF; Vascade S1000) with a manufacturer-specified normal GVD coefficient of $\beta_{2,\mathrm{DCF}} = 48$~$\mathrm{ps^2/km}$ for GVD compensation. The total fiber link is 22.3~m in length.

We stabilize the optical path length of the fiber link by detecting the time-dependent interference pattern between the residual pump component of the sync signal and the second CW pump. This signal is detected with a photodiode and is split into two signals. One is monitored on the oscilloscope while the other is sent to a commercial proportional-integral controller (New Focus LB1005) to generate an error signal, defined as the difference between the detected signal and the internal reference level. The controller actuates on a fiber stretcher in the link according to the error signal and stabilizes the interference signal.

The setup for autocorrelation measurement is summarized with a simple schematic in Supplementary Fig.~S2. It only shows a portion of the setup that is relevant to autocorrelation measurement, and the rest is identical to the schematic of Supplementary Fig.~S1. We add a polarization controller to each input arm of the 50/50 coupler, which allows independent control over the individual polarization states of the two comb outputs. A fiber-coupled variable delay line is added to one arm to control the relative temporal delay between the two combs. The comb signals are then combined and amplified with an ultrashort optical pulse amplifier to an average power of about 30~mW. A length of DCF is added after the amplifier to compensate for the GVD of the fiber links. By ensuring that the input arms of the 50/50 coupler are similar in length, we are able to compensate for the GVD of the both arms with a single length of DCF. Each trace of Fig. 2a and b in the main article corresponds to a fixed delay between the two arms. The delay is varied in steps of 0.075~mm over a total range of 1.5~mm, covering one pulse period (5 ps) of a 200~GHz pulse train.

\subparagraph*{Beatnote measurement and synchronization range. }
After accessing the single cavity soliton regime, the waveform generator supplies a constant DC offset to the microheaters to sustain the state. Tuning this offset voltage alters the resonance frequency detuning from the pump~\cite{joshi_thermally_2016,carmon_dynamical_2004}, and hence the characteristics of the circulating cavity solitons~\cite{leo_temporal_2010}. This is the basis of the measurements in Figs.~\ref{fig:fig3}c and e. For Fig.~\ref{fig:fig3}c, we systematically tune the heater voltage of the slave microresonator in steps of 1~mV, and for each voltage, acquire the beatnote trace. We leave the master microresonator unaltered. The fundamental beatnote is extracted from each trace. For Fig.~\ref{fig:fig3}e, this procedure is repeated multiple times for a range of different transmission coefficients of the coupling fiber link. The coupling transmission is varied by inserting different absorptive filters. Additionally, we insert a 99/1 coupler in the fiber link where the 1~\% tap allows us to track the coupling strength without introducing much loss. For each transmission value, we extract the frequency of the first beatnote detected on either side of the synchronization range. For more details, see Supplementary Fig.~S4.

For Fig.~\ref{fig:fig3}f, we replace the absorptive filters with a combination of a half-wave plate and an in-line polarizer, which enables continuous variation of the transmission. We first set the heater voltage of the slave microresonator such that the fundamental beatnote is observed at approximately 2~MHz. The beatnotes are detected as we manually rotate the half-waveplate.

\subparagraph*{Calibration of microheater voltage. }
In order to measure the calibration factor between the heater voltage and the corresponding thermal shift of the resonances, we externally modulate the frequency of the CW pump laser with a triangular waveform. The total frequency shift in this case is 7.68~GHz. The DC heater voltage is adjusted so that the nearest resonance appears within this range. We measure resonance shift as the voltage is further varied. From this procedure, we calculate the calibration factor to be 35.6~MHz/mV. Therefore, the synchronization range of 11~mV in Fig.~\ref{fig:fig3}c corresponds to approximately 400~MHz thermal shift of resonance.

\subparagraph*{Numerical simulation. }
Our theoretical model is based on a set of Lugiato-Lefever equations~\cite{coen_modeling_2013,chembo_spatiotemporal_2013,wabnitz_suppression_1993,lugiato_spatial_1987}, which are numerically integrated by employing the split-step Fourier method with a step size corresponding to one roundtrip time of the master resonator. The span of the temporal grid is also chosen to be the roundtrip time in order to satisfy the periodic boundary condition of the resonator. The grid is discretized into $n = 1,024$ points which are sufficient to describe the full spectral widths of our combs. For more details on the model and parameter values, see Supplementary Information.

\bigskip

\section*{Acknowledgements}

\noindent This work was supported by Air Force Office of Scientific Research (AFOSR) (grant FA9550-15-1-0303), National Science Foundation (NSF) (grants CCF-1640108 and PHY-1707918), and Semiconductor Research Corporation (SRC) (grant SRS 2016-EP-2693-A). We also thank K. Bergman and R. Polster for kindly lending us the high-resolution optical spectrum analyzer and the autocorrelator.

\section*{Author Contributions}

\noindent J.K.J and A.K. performed experiment. J.K.J carried out theoretical analysis and numerical simulation, and wrote the manuscript with inputs from all authors. J.K.J., A.K., Y.O. and A.L.G. contributed to the interpretation of data. X.J. fabricated the devices under the supervision of M.L. A.L.G. supervised the overall project.

\section*{Additional information}

\noindent The authors declare no competing financial interests. Correspondence and requests for materials should be
addressed to A.L.G.

\end{document}